# PrisCrawler: A Relevance Based Crawler for Automated Data Classification from Bulletin Board

Pu Yang, Jun Guo, and Weiran Xu
*School of Information and Telecommunication Engineering*
*Beijing University of Posts and Telecommunications,*
*Beijing, 100876, China*
*yangpu1015@gmail.com, {guojun, xuweiran}@bupt.edu.cn*

**Abstract**

*Nowadays people realize that it is difficult to find information simply and quickly on the bulletin boards. In order to solve this problem, people propose the concept of bulletin board search engine. This paper describes the priscrawler system, a subsystem of the bulletin board search engine, which can automatically crawl and add the relevance to the classified attachments of the bulletin board. Priscrawler utilizes Attachrank algorithm to generate the relevance between webpages and attachments and then turns bulletin board into clear classified and associated databases, making the search for attachments greatly simplified. Moreover, it can effectively reduce the complexity of pretreatment subsystem and retrieval subsystem and improve the search precision. We provide experimental results to demonstrate the efficacy of the priscrawler.*

## 1. Introduction

In the twenty-first century, the forms of the social work are carrying on the earth-shaking changes: from the previous using chalk to write on the blackboard to make notices to then using the papers posted on the board to inform people, and now to using no paper on the bulletin board. The bulletin board on the web has many advantages, for example, easy-to-update, large amount of storage, need no paper, more to public and so on. These advantages make an increasing number of schools, businesses, etc. to set up bulletin board system in order to improve the efficiency of study and work.

Because of the bulletin board easy to update and large storage, people are merged in a lot of information. Even if some of the existing bulletin board can show the notices with classification, but it is also far from satisfactory, because studies have shown that almost 90 percent of information on the bulletin board exists in the attachments, such as doc, txt, xls and so on. All of these information in the attachments can not be obtained by people via notices with classification. As a result, the bulletin board search engine is a very critical problem. Crawler as a data-collecting subsystem of bulletin board search engine, naturally be concerned by lots of scholars.

In such particular environment--the bulletin board, the attachment is different from the page; it does not contain any hyperlink information and is in the chaos of existence on the bulletin board. Therefore, the crawler's ability of handling attachments directly determines the success or failure of the bulletin board search engine. The current crawler subsystems [1][2][3][4] are designed for webpages, thus it can not solve the non-relevance and chaos of the attachments on the bulletin board. At present, there is no effective method of building crawler subsystem of bulletin board search engine.

In this paper, we propose a novel solution to solve the non-relevance and chaos of the attachments. Research made by us shows that users who access to the bulletin board are actually to find the attachments not webpages, because the main information on the bulletin board is in attachments. And the webpages on the bulletin board always include titles, times, the information about what the attachments are. Thus the attachments and the webpages containing them have an extremely strong relevance. In order to solve the non-relevance of the attachments, we introduce the Attachrank algorithm, which can clearly show the hidden relevance between attachments and webpages and give the attachments hyperlink information. In other words, Attachrank algorithm which is similar to pagerank algorithm is used for revealing the "importance" of the attachment given by the other attachments via webpages' hyperlinks. What is more, we classify the attachments according to their different types in order to solve the chaos of the attachments and produce clear classified and associated databases.

Priscrawler is a specialized crawler that uses our novel solution to turn bulletin board into the special databases. Such databases enable construction of bulletin board search engine.

The rest of this paper is organized as follows. Section 2 outlines the framework of the priscrawler and describes various strategies for building the system. Section 3 describes the Attachrank algorithm. Section 4 presents experimental results. Section 5 present related works and section 6 concludes the paper.

## 2. Framework of priscrawler

### 2.1. Loop of priscrawler

The basic actions of priscrawler are similar to those of other traditional crawlers [2][3]. In Figure 1 the flowchart indicates the typical crawler loop, consisting of URL selection, page retrieval, and page processing to extract links. Note that traditional crawlers do not deal with the attachments, actually just discard them. However, as shown In Figure 2, priscrawler's execution sequence contains additional novel steps for attachments in which attachments are classified and attachranked. Specifically, priscrawler performs the following sequence of actions for each page:
● **Step 1** Retrieve and store the page of bulletin board to build a webpage database and prepare for the next step. (Retrieve page and Store page)
● **Step 2** Scan hyperlink contained by page and send hyperlink to be analyzed. (Scan link)
● **Step 3** Analyze the hyperlink to check if the hyperlink point to an attachment. (Attachment)
● **Step 4** If the hyperlink point to an attachment, the attachment is classified via suffix of the hyperlink. (Classify)
● **Step 5** Use Attachrank algorithm to add relevance to attachment. The Attachrank algorithm is introduced in Section 3. (Attachrank)
● **Step 6** Store the attachment with relevance information added on the head of it. (Store attachment)
● **Step 7** If the hyperlink doesn't point to an attachment, analyze and check if the hyperlink point to another page. (Page)
● **Step 8** If the hyperlink points to a page, store this hyperlink. (Store hyperlink)
● **Step 9** If the hyperlink doesn't point to a page, just throw away this hyperlink. (Throw away)
● **Step 10** Analyze and check if the all of the hyperlinks contained by current page are dealed with. If not, Steps 2 is executed again. (Done with page)
● **Step 11** If all of the hyperlinks are handled, analyze and check if the all of the stored hyperlinks are run out. If not, Steps 1 is executed again. (Done with hyperlink)

All of the steps are executed repeatedly, until all of the hyperlinks are treated.

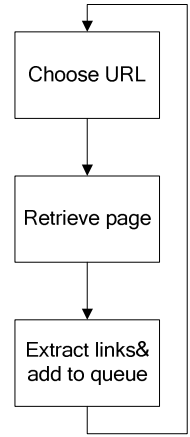

**Figure** 1**. Traditional crawler loop**

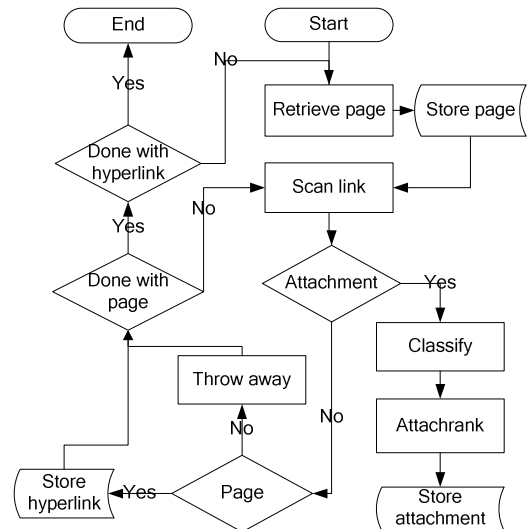

**Figure** 2**. Priscrawler loop**

### 2.2. Architecture of priscrawler

Figure 3 illustrates the complete architecture of the priscrawler. It includes eight basic functional modules. When starting up the crawler, Link Provider transports the hyperlink pointing to webpage from Link Storager to Page Retriever and Attachrank Producer. The Page Retriever retrieves the webpage and sends it to Link Scanner. The Link Scanner scans and finds one of the hyperlinks of the current page in order to be analyzed by Attachment Classifier. The Attachment Classifier analyzes the current hyperlink and decides how the current hyperlink should be treated. If the current hyperlink points to another webpage, it should be sent to the Link Storager; if the current hyperlink points to attachment, it should be sent to Attachrank Producer. The Link Storager saves the current hyperlink which is sent by the Attachment Classifier. The Attachrank Producer receives the hyperlink sent by the Link Provider which points to the current page and is different from the current hyperlink which points to the attachment and is sent by the Attachment Classifier. And then the Attachrank Producer uses both hyperlinks to generate the relevance information. Relevance Writer writes the relevance information to the head of the attachment. Attachment Storager saves the classified attachment which concludes the relevance information.

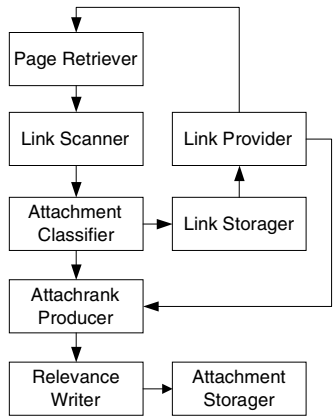

**Figure** 3. **The architecture of priscrawler**

## 3. Attachrank algorithm

First let us review the PageRank [5][6][7][8][9][10]. PageRank is defined as follows:

We assume page A has pages T1...Tn which point to it (i.e., are citations). The parameter d is a damping factor which can be set between 0 and 1. We usually set d to 0.85. Also C(A) is defined as the number of links going out of page A. The PageRank of a page A is given as follows [5]:

$$PR(A) = (1-d) + d(\frac{PR(T1)}{C(T1)} + ... + \frac{PR(Tn)}{C(Tn)}) \quad (1)$$

Note that the PageRanks form a probability distribution over web pages, so the sum of all web pages' PageRanks will be one.

Then we review the meaning and the worth behind the PageRank [9]. PageRank is a method of measuring a page's "importance." To examine the worth of PageRank, we need to first look at its premise, and how accurate it is. Basically PageRank [5] says: 1. If a page links to another page, it is casting a vote, which indicates that the other page is good. 2. If lots of pages link to a page, then it has more votes and its worth should be higher. The basic implication here is: People only link to pages they think are good. But the premise may not always be true. A few of the reasons people link to pages other than ones they think are good are:
1) Reciprocal links – "Link to me and I'll link to you."
2) Link Requirements – "Using our script requires you to put a link to our page." or "We'll give you an award solely because you link to our page." 3) Friends and Family – "This is my friend Pete's site." or "My mum's site is here, my dad's site is here. My dog's site is here."

All of above scenarios happen on the Internet and affect the precision of PageRank. Research made by us shows that they don't take place on the bulletin board. We checked 50 different bulletin boards of different companies and schools which were randomly selected. We find that almost all of the bulletin boards are carefully managed and only used for informing people, so they contain few ads and almost all of the links point to webpages on the bulletin boards or attachments. Another result shows that users who access to the bulletin board are actually to find the attachments by first finding webpages which conclude them, because the main information on the bulletin board is in attachments. And the webpages on the bulletin board always include titles, times, the information about what attachments are. Thus the attachments and the webpages containing them have an extremely strong relevance.

Based on the above two results, we introduce the Attachrank algorithm. Because the hyperlinks in webpages of the bulletin board point to webpages on the bulletin boards or attachments, all of the pages casting a vote each other are effective. Because users who access to the bulletin board are actually to find the attachments by first finding webpages which conclude them, the webpage and the attachments contained in this webpage have the same value and the same pagerank. We assume page A has attachments M1...Mn. AR(Mi) indicates the attachrank of attachment Mi, i is a positive integer and can be selected from 1 to n. Therefore, we define Attachrank of an attachment as follows [5]:

$$PR(A) = (1-d) + d(\frac{PR(T1)}{C(T1)} + ... + \frac{PR(Tn)}{C(Tn)}) \quad (2)$$

$$AR(Mi) = PR(A) \quad (3)$$

(2) and (3) also equal to

$$PR(A) = AR(M1) = ... = AR(Mn) \qquad (4)$$

## 4. Experimental results

In this Section, we present the experimental results of the proposed system and some comparisons with the system without using the Attachrank, in terms of the precision.

### 4.1. Experiment setup

In our experiment, we randomly selected the office bulletin board and student bulletin board of Beijing University of Posts and Telecommunications (BUPT) as research targets as shown in Table 1. In order to set up a consistent data collection for further evaluation and comparison, we first used normal crawler to collect about 60,000 webpages and attachments from bulletin boards in Table 1. Second, we mirrored these webpages and attachments as a bulletin board. Third, we used priscrawler based on Attachrank algorithm to crawl from the mirrored bulletin board to build the clear classified and associated databases. Finally, we had the two databases with the same webpages and the same attachments: the one is normal structure database; the other is clear classified and associated database. We used these two databases for further comparison.

**Table 1. Research bulletin boards**

| Name | URL |
|---|---|
| office bulletin board of BUPT | http://buptoa.bupt.edu.cn/ |
| student bulletin board of BUPT | http://buptoa.bupt.edu.cn/student_broad.nsf |

In order to eliminate the effect of different pretreatment subsystem and retrieval subsystem to comparing result, we utilized the same pretreatment subsystem and retrieval subsystem for different two databases. We use lemur [11] to build the retrieval subsystem.

### 4.2. Evaluation of priscrawler

In this section, we evaluate the performance of priscrawler. We used the same subsystem and retrieval subsystem to deal with the different databases. In order to get the accurate results, we keep all environments constant: the same PC, the same subsystem and so on.

In order to eliminate the effect of the particular data to comparing result, we first used the normal database to be handled by pretreatment subsystem. And then we used the classified and associated database which is generated by priscrawler to be dealt with by pretreatment subsystem in the same condition. Third, we repeated the above two steps six times. Finally, we calculated the average handling time of the pretreatment subsystem. The handling time of each round is listed in Table 2.

**Table 2. The pretreatment subsystem's handling time of each round**

| Round | Normal Database (minutes) | Classified and Associated Database |
|---|---|---|
| 1 | 70 | 66 |
| 2 | 75 | 73 |
| 3 | 80 | 57 |
| 4 | 83 | 59 |
| 5 | 66 | 68 |
| 6 | 74 | 58 |

The average handling time of the pretreatment subsystem treating normal database is 74.7 minutes; and the average handling time of the pretreatment subsystem treating classified and associated database is 63.5 minutes. With all environments constant, the average system complexity can be illustrated by the average handling time. Thus, the average decrease proportion of complexity of pretreatment subsystem is 15%, which is shown in Figure 4.

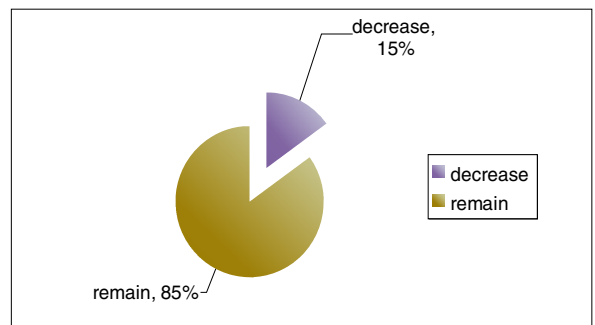

**Figure 4. The average decrease proportion of complexity of pretreatment subsystem**

With the same conditions, the reason that leads to this good performance is the clear classified and associated databases, which make the parsers of the pretreatment subsystem do less duplication of work with the classified and associated database.

We next evaluated the performance of classified and associated database to retrieval subsystem. In order to eliminate the effect of the pretreatment subsystem to search time, we first used pretreatment subsystem to handle the two different databases and we used the two handled databases do our next experiments. Moreover, we chose one query to be searched 10 times in normal database and in classified and associated database. In addition, we calculated the average searching times. The searching times are listed in Table 3.

**Table 3. The retrieval subsystem's searching time of each time**

| Time | Normal Database (seconds) | Classified and Associated Database |
|---|---|---|
| 1 | 0.09 | 0.05 |
| 2 | 0.08 | 0.06 |
| 3 | 0.10 | 0.07 |
| 4 | 0.13 | 0.09 |
| 5 | 0.15 | 0.08 |
| 6 | 0.08 | 0.06 |
| 7 | 0.11 | 0.10 |
| 8 | 0.09 | 0.09 |
| 9 | 0.12 | 0.10 |
| 10 | 0.07 | 0.12 |

The average searching time of the retrieval subsystem treating normal database is 0.102 seconds; and the average searching time of the retrieval subsystem treating classified and associated database is 0.082 seconds. With all environments constant, the average system complexity can be illustrated by the average searching time. Thus, the average decrease proportion of complexity of retrieval subsystem is 20%, which is shown in Figure 5.

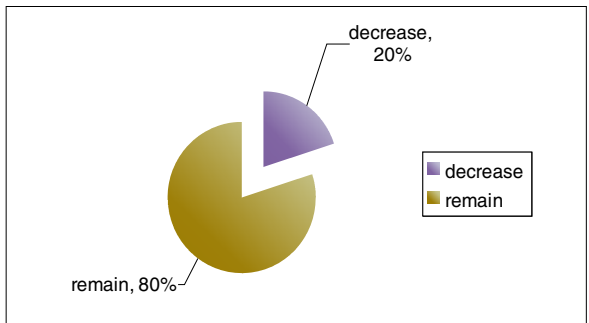

**Figure 5. The average decrease proportion of complexity of retrieval subsystem**

With the same conditions, the use of the Attachrank algorithm makes retrieval subsystem does less interaction between the retrieval subsystem and the data and less repeat operations.

What is more, we evaluated the performance of the search precision before and after using attachrank algorithm.

In order to decrease the effect of particular query to comparing result, we randomly chose 6 groups of test queries and each group had 50 test queries. We used 50 queries of each group to test normal database and the classified and associated database and calculated the average precision of each group. The values of the average precisions of each group are listed in Table 4.

**Table 4. The average precisions of each group**

| Group | Before | After |
|---|---|---|
| 1 | 0.627 | 0.844 |
| 2 | 0.610 | 0.873 |
| 3 | 0.749 | 0.850 |
| 4 | 0.778 | 0.867 |
| 5 | 0.804 | 0.909 |
| 6 | 0.512 | 0.757 |

The comparison of search precision before and after using Attachrank algorithm is shown in Figure 6. From it, we can see that in the use of the Attachrank algorithm, the average search precision of the bulletin board search engine is 85% and without using it, the average search precision is 68%.

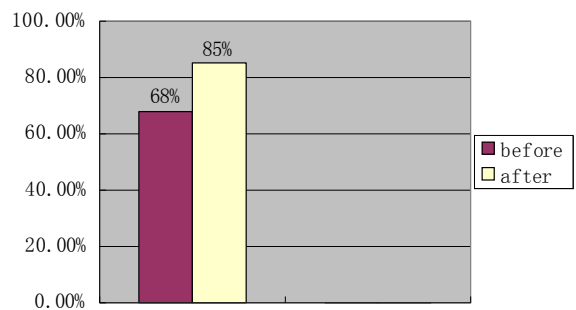

**Figure 6. The comparison of the average search precision before and after using attachrank algorithm**

The reason that leads to precision improvement is the use of the Attachrank algorithm. The traditional methods search attachments via page. These methods bring inprecision in the search results. On the contrary, when the crawler is based on the attachrank algorithm, it brings relevance in attachments. Thus all of the attachments on the bulletin board build up a new virtual net where the attachments are connected by attachranks. We use attachrank to modify the search result. Such way can no doubt improve the search precision.

## 5. Related work

Information extraction from the web is a well-studied problem. But most of the crawler subsystems [1][2][3][4] are designed for webpages, thus it can not solve the non-relevance and chaos of the attachments on the bulletin board. Therefore, the search engines based on these crawlers search attachments by searching the pages containing the attachments. Instead, our information extraction technique uses Attachrank algorithm, which brings relevance in attachments. Thus all of the attachments on the bulletin board build up a new virtual net where the attachments are connected by attachranks. The other difference between our system and [1][2][3][4] is that we also classify the attachments in order to provide the clear classified and associated databases.

## 6. Conclusion

This paper describes the priscrawler system, a subsystem of the bulletin board search engine, which can automatically crawl and add the relevance to the classified attachments of the bulletin board. Priscrawler utilizes Attachrank algorithm to bring forth the relevance between webpages and attachments and turns bulletin board into clear classified and associated databases, making the search of attachments greatly simplified just like the search for the webpages. Thus it can effectively reduce the system complexity of pretreatment subsystem and retrieval subsystem and improve the search precision.